\documentclass[10pt,a4paper]{article} 

\usepackage[,margin=1.1in]{geometry}
\usepackage[latin1]{inputenc}
\usepackage{amsmath}
\numberwithin{equation}{section}
\usepackage{amsfonts}
\usepackage{amssymb}
\usepackage{amsthm}
\usepackage{enumitem}
\usepackage{tikz}
\usetikzlibrary{shapes, calc, chains}
\usepackage{algorithm}
\usepackage{algorithmic}
\usepackage{listings}
\usepackage{color}
\usepackage{graphicx}
\usepackage{float}
\usepackage{caption}
\usepackage{xcolor}
\usepackage[font=footnotesize,labelfont=bf]{caption} 
\usepackage[multiple]{footmisc}
\usepackage{titlesec}
\usepackage[many]{tcolorbox} 
\usepackage{hyperref}
\usepackage{footnote}
\makesavenoteenv{tabular}
\makesavenoteenv{table}

\titleformat*{\section}{\large\bfseries}
\titleformat*{\subsection}{\normalsize\bfseries}
\titleformat*{\subsubsection}{\normalsize\bfseries}
\titleformat*{\paragraph}{\normalsize\bfseries}
\titleformat*{\subparagraph}{\normalsize\bfseries}

\newtheorem*{theorem*}{Theorem}
\newtheorem{theorem}{Theorem}

\newtheorem*{lemma*}{Lemma}
\newtheorem{conj}[theorem]{Conjecture}

\theoremstyle{definition}
\newtheorem{definition}{Definition}[section]

\theoremstyle{remark}

\newcommand{\isep}{\mathrel{{.}\,{.}}\nobreak}

\newtcolorbox{titledbox}[1]{
    tikznode boxed title,
    enhanced,
    arc=0mm,
    interior style={white},
    attach boxed title to top center= {yshift=-\tcboxedtitleheight/3},
    fonttitle=\large\bfseries,
    colbacktitle=lightgray,
    coltitle=black,
    boxed title style={size=normal,colframe=white,boxrule=2pt},
    title={#1}
}

\newtcolorbox{titledboxinside}[1]{
	tikznode boxed title,
	enhanced,
	arc=1mm,
	interior style={white},
	attach boxed title to top center= {yshift=-\tcboxedtitleheight/5},
	fonttitle=\large\bfseries,
	colbacktitle=white,
	coltitle=black,
	boxrule=1pt,
	boxed title style={size=small,colframe=black, boxrule=1pt},
	title={#1}
}

\DeclareRobustCommand{\heapName}{Logarithmic Funnel} 

\title{\LARGE{The \heapName{}:\\An Efficient Priority Queue For Extremely Large Sets}}
\author{Christian Loeffeld\footnote{Email: c.loeffeld@proton.me}}

\begin{document}
\maketitle 
\begin{abstract}	
\noindent For many data-processing applications, a comprehensive set of efficient operations for the management of priority values is required. Indexed priority queues are particularly promising to satisfy this requirement by design. In this work, we report the design and analysis of an efficient indexed priority queue with a comprehensive set of operations. In particular, $\mathtt{insert}$, $\mathtt{delete}$ and $\mathtt{decrease}$ all run in expected $O(\log^{*}{n})$ time, while $\mathtt{increase}$ is conjectured by means of Monte Carlo simulations to run in expected $O(\log\log{n})$ time. The space complexity as well as the time complexity for the construction of the empty heap data structure is $O(n)$. For certain massive computational problems, such as specific analyses of extremely large graphs and (chemical) simulations, this heap system may exhibit considerable utility.
\end{abstract}

\section{Introduction}

A heap data structure maintains a specific partial ordering of a collection of comparable types. The heap is one of the most ubiquitously utilized data structures in computer science. Priority queues are employed in a broad range of important applications, from simulations \cite{Gibson, Ajay}, to scheduling \cite{Hou, Laksmi}, and graph analysis \cite{BackToBasics}. A broad survey into the variety of forays into heap research revealed a large number of ideas and approaches that have shaped the development of priority queues and beyond \cite{Brodal_Survey}. \\

In 1964, Williams \cite{Williams} was the first to introduce a sorting algorithm based on the notion of an array-based heap. An example of an efficient priority queue in practice is based on \textit{van Emde Boas Trees} \cite{EmdeBoas} and widely used in network routers. Various priority queues have been devised that exhibit optimal running times for common heap operations. Those ideas transformed the ability to analyze complex graph problems. Among them, the \textit{Pairing-Heap} \cite{Fredman-pairing}, the  \textit{Fibonacci heap} \cite{Fredman-fibonacci}, \textit{ Brodal Queues} \cite{Brodal}, and the \textit{Rank-Pairing Heap} \cite{Haeupler}. Another group of efficient data structures for shortest-path algorithms turn out to be so-called multi-level bucket heaps \cite{Cherkassky1,Cherkassky2}. For applications  employing monotone sequences and involving shortest-path algorithms, especially efficient monotone priority queues have been designed \cite{Raman, Thorup}. \\

In this work, we present the design and analysis of an indexed priority queue, named the \heapName{}. This priority queue maintains uniquely identifiable elements that contain priority values in heap-order and supports comprehensive functionality for efficient modification of the priorities. Additionally, it also enables efficient deletion of elements from the priority queue. We will demonstrate runtime and space characteristics of this heap system in the realization of a max-heap, which besides the common heap operations, namely $\mathtt{insert, find\textrm{-}max, delete\textrm{-}max}$, also facilitates the specific heap operations, $\mathtt{increase, decrease}$ and $\mathtt{delete}$ on its elements $[id:p]$, where $id$ represents an immutable identifier and $p$,the priority, a comparable and mutable type.  For a summary of the runtime behavior, see Table \ref{fig:complexity-summary} below.\\

\begin{table}[ht!]
\centering
	\scalebox{0.85}{
			\begin{tabular}{lllc}
			\hline
			Name  & Operation  &  Time Complexity & Section  \\
			\hline
			\hline
			$\mathtt{make\textrm{-}heap(\lambda, \textit{n})}$   & Make empty ($\lambda$+1)-level heap for $n$ elements.  &  $O(n)$  &  \ref{make-heap}     \\
			$\mathtt{find\textrm{-}max()}$ & Access element with highest priority. & $O(1)$  &   \ref{find-max}     \\
			
			$\mathtt{delete\textrm{-}max()}$ & Delete element with the highest priority. &   $O(\log{n})$ & \ref{delete-max}       \\
			
			$\mathtt{insert(id,p)}$ & Insert element $id$ with priority $p$. &   $O(\log^{*}{n})$ \footnote[2]{Expected running time.}   &  \ref{insert}    \\
			
			$\mathtt{delete(id)}$ & Delete the element $id$.  &  $O(\log^{*}{n})$ \footnotemark[2]    &   \ref{delete}   \\
					
			$\mathtt{decrease(id,p')}$ & Decrease priority from $p$ to $p'$ of element $id$. &  $O(\log^{*}{n})$ \footnotemark[2]  &   \ref{decrease}      \\
			
			$\mathtt{increase(id,p')}$  & Increase priority from $p$ to $p'$ of element $id$. &   $O(\log\log{n})$ \footnotemark[2]\footnote[3]{Projected upper-bound of expected running time - determined using Monte Carlo simulations. See Section \ref{increase}.}   &  \ref{increase}     \\
			\hline
		\end{tabular}
}
\caption{Overview of heap system operations and associated time complexities on elements with unique immutable identifiers $id$ and mutable, potentially degenerate priorities $p$. The expression $\log^{*}{n}$ denotes the iterated logarithm (see Section \ref{level-depth} for more details).\label{fig:complexity-summary}}
\end{table}

Beyond the present overview, this report contains three additional sections. Basic properties and structural composition of the \heapName{} are detailed in Section \ref{structure}. The functional facilities supported by this heap system are analyzed and its associated computational complexities are established in Section \ref{analysis}. In Section \ref{conclusion}, we then conclude with some remarks, allude to some open questions, and mention a few potential extensions to this heap system.
\section{Structure and Composition of the \heapName{} $H_{\lambda}$}\label{structure}

The \heapName{} is an indexed priority queue. It is a recursive and composite data type built from two families of specific composite heap types, namely the base heap type $H_{0}$ and the inductive heap types $H_{\ell}$ with $ 0 < \ell \le \lambda$. The recursive nature and data type composition of this heap system are graphically illustrated in Figure \ref{fig:structure}. A \heapName{} with $\lambda$ inductive levels is denoted as $H_{\lambda}$, and is constructed from multiple instances of all $\lambda$ distinct heap types $H_\ell$ where $\ell \in [1\isep\lambda]$, and multiple instances of base heaps $H_{0}$.\\

\begin{figure}[h!]
	\centering
	\tikzset{
		treenode/.style = {align=center, inner sep=0pt, text centered, font=\sffamily},
		hashNode/.style = {align=center, rectangle, draw=black, minimum height=10mm, minimum width=10mm, draw},
		commonNode/.style={rectangle, draw=black, draw},
		triangle/.style={isosceles triangle,  draw,shape border rotate=90, minimum height=7mm, minimum width=9mm},
	}
\scalebox{0.7}{
	\begin{tcolorbox}
		
		\begin{tikzpicture}[level/.style={scale=1.5,sibling distance = 4cm/#1,level distance = 0.5cm}] 
		\node[hashNode] at (11.2, 0.9) {$\mathbb{T}$};
		\end{tikzpicture}

		\begin{titledbox}{$H_{0}$}
		\begin{tikzpicture}[level/.style={scale=1.5,sibling distance = 4cm/#1,level distance = 0.5cm}] 
		\node[hashNode] at (3.2, 0) {$\mathbb{B}$$H_{1}$};
		\node [commonNode] at (9., -0.5) {$id:p$} 
		child{node [commonNode]  {$id:p$} 
			child{node [commonNode]  {$id:p$} } 
			child{node [commonNode]  {$id:p$} }                            
		}
		child{ node [commonNode]  {$id:p$}
			child{node [commonNode]  {} }
			child{node [commonNode]  {} }
		};
		\end{tikzpicture}
	\end{titledbox}

	\begin{titledbox}{$H_{\ell}$}
	\begin{tikzpicture}[scale=0.4]
	\node[hashNode] at (3.2, 0.9) {$\mathbb{B}$$H_{\ell+1}$};
	\node[hashNode] at (8.2, 0.9) {$\mathtt{\textit{\textit{\textit{heap\_max}}}}$}; 
	\end{tikzpicture}
	
		\begin{titledboxinside}{$\mathbb{M}$}
			\begin{tikzpicture}[scale=1.16, level/.style={sibling distance=6cm/#1, level distance =.92cm}]	
			\node [commonNode]  {$\mathbb{P}$$H_{\ell-1}$}
			child{ node [commonNode] {$\mathbb{P}$$H_{\ell-1}$} 
				child{ node [commonNode] {$\mathbb{P}$$H_{\ell-1}$}
					child{ node [triangle] {}} 	
				}
				child{ node [commonNode]  {$\mathbb{P}$$H_{\ell-1}$}
					child{node [triangle] {}} 	
				}                            
			}
			child{ node [commonNode]  {$\mathbb{P}$$H_{\ell-1}$}
				child{ node [commonNode]  {$\mathbb{P}$$H_{\ell-1}$}
					child{node[triangle] {}}   	
				}
				child{ node [commonNode]  {$\mathbb{P}$$H_{\ell-1}$}
					child{node[triangle] {}}  
				}
			};
			\end{tikzpicture}
		\end{titledboxinside}
	
	\end{titledbox}
	\vspace{1em}
	\underline{\textbf{LEGEND}}
	\begin{itemize}[ label={}, leftmargin=*]
		\item $\mathbb{T}$: A hash table that manages $[id : address]$-pairs, denoted as \textit{location\_table}.
		\item $id$: Unique and immutable identifier of an element.
		\item $p$: Comparable and mutable priority of an element.
		\item  $\mathbb{M}$: Array-based binary heap managing pointers to $H_{\ell-1}$ heaps. Denoted as \textit{metaheap}.
		\item $\mathbb{P}$$H_{\ell-1}$: Pointer to a $H_{\ell-1}$, i.e. child heap.
		\item $\mathbb{B}$$H_{\ell+1}$: Backpointer to a specific position in the \textit{metaheap} of the parent $H_{\ell+1}$ heap.
	\end{itemize}
	\end{tcolorbox}
}
	\caption {Graphical representation of the composition and structural setup of the \heapName{}. The heap system is characterized by its recursive depth $\lambda + 1$, and constructed from a large number of instances of two structurally distinct heap families, namely heaps of type $H_{\ell}$ where $\ell \in [1\isep \lambda]$, and the base heap type $H_{0}$. It contains a global hash table $\mathbb{T}$ to store routing paths from level $\lambda$ to the base level.\\\textbf{Top ($H_{0}$):} The base heap $H_{0}$ consists of a back-pointer $\mathbb{B}$ to its parent $H_{1}$ heap and an implicit binary max-heap that maintains a specific set of elements $id:p$, where \textit{id} is a unique, immutable identifier and $p$ is a mutable and comparable value.\\\textbf{Bottom ($H_{\ell}$):} Each inductive heap $H_\ell$ consists of an implicit binary max-heap, denoted as \textit{metaheap} $\mathbb{M}$, a back-pointer $\mathbb{B}$ to the parent $H_{\ell+1}$ heap, and a variable $\mathtt{\textit{heap\_max}}$ to store the max priority value of the particular sub-heap system. The elements of each \textit{metaheap} $\mathbb{M}$ of an inductive heap $H_\ell$ are pointers $\mathbb{P}$ to heaps of type $H_{\ell-1}$.		
	\label{fig:structure}}
\end{figure}

The heap system contains a single global hash table, the \textit{location\_table} $\mathbb{T}$, to store a specific set of $[id: [heap-addresses]]$-bundle. The table $\mathbb{T}$ essentially acts as a routing table facilitating a unique path down the system to the particular base heap $H_{0}$ that stores the required element that is associated with the unique identifier $id$.

\subsection{Base Heap Type $H_{0}$}\label{base-heaps}

Structurally, base heaps $H_{0}$ are a small composite data structures that in addition to core heap functionalities, also support naive operations for decrease and increase of the priority value of an element, as well as deletion of an element. Base heaps $H_{0}$ consist of two distinct data types, (i) a backpointer to its parent inductive heap, a heap of type $H_{1}$, and (ii) an array-based, implicit max-heap that maintains a very small set of constant-size elements of the form $id:p$, where $id$ is a unique and immutable identifier and $p$ is a mutable and comparable priority value.\\

The data elements of the heap system $H_{\lambda}$ are exclusively managed by $H_{0}$ heaps, i.e. all element operations occur only on the base-level. The inductive level functions of the heap system only serve to direct the requested action on the element to the particular $H_{0}$ base heap hosting the specific element. \\

See Section \ref{level-depth} and Figure \ref{fig:small-base-heaps} for more details on base heap sizes.
An exciting recent results seemingly supporting this particular design choice was presented by Mankowitz, Michi, Zhernov \textit{et al.}\cite{Mankowitz}. They discovered highly efficient, optimized algorithms to sort short sequences of elements. Benchmarks using these sort implementations in the LLVM C\texttt{++} library \cite{Gelmi} revealed improvements of up to 70\% for sequences of a length of five for uint32, uint64 and float on specific architectures \cite{Mankowitz}.

\subsection{Inductive Heap Types $H_\ell$}\label{inductive-heaps}

The composition of the inductive heap types $H_\ell$ is being depicted in the bottom half of Fig. \ref{fig:structure}. Each heap of type $H_\ell$ consists of an array-based binary max-heap denoted as $\mathtt{\alpha}$ $\mathbb{M}$, a backpointer $\mathbb{B}$ to the \textit{metaheap} of its parent heap, a $H_{\ell+1}$ heap, and a variable $\mathtt{\textit{\textit{\textit{heap\_max}}}}$ to keep track of the local max-priority element. \\

The elements of the \textit{metaheap} of an inductive heap $H_\ell$ are pointers $\mathbb{P}$ to heaps of type $H_{\ell-1}$. The \textit{metaheap} implicitly maintains in heap-order the max-priority elements of its $k_{\ell}$ bidirectionally linked heaps of type $H_{\ell-1}$. The bidirectionality of the heap linkage is required in order to compute in constant time the position in the \textit{metaheap} where a heap-property violation may have occurred as the result of an element deletion or the modification of its priority value.

\subsection{The Depth \textbf{$\lambda_{\alpha}$} of the Heap System $H_\lambda$}\label{level-depth}

The time and space computational complexities associated with the operations inscribed to this heap system are defined by its level depth $\lambda$, or equivalently the shortest number of steps from level $\lambda$ to the base which maintain all data elements. The quantity $\lambda$ is determined by employing a small variation to the standard definition of the iterated logarithm \cite{itlog}.

\begin{definition}[The Generalized Iterated Logarithm]
	Let $\alpha\in\mathbb{R}^{+}_{1}$, then we define the generalized iterated logarithm as follows,
	\begin{equation} \label{def:log-star}
		\log^{\alpha\ast} n := 
		\begin{cases} 
			0 & \text{if } n \leq \alpha \\
			1 + \log^{\alpha\ast}(\log n)  & \text{if } n > \alpha
		\end{cases}
	\end{equation} 
\end{definition}

This expression defines the number of times the logarithm must be iteratively applied before the result is less than or equal to a small predefined constant value, $\alpha$. Definition \ref{def:log-star} is introduced as extension to the common iterated logarithm, where $\alpha = 1$, and in order to guide the construction of sufficiently small base heaps $H_{0}$, and consequently to enforce associated base heap operations with expected constant time complexities.\\

For convenience, we denote the level depth $\lambda_{\alpha}$ of the \heapName{} using Definition \ref{def:log-star} as follows,
\begin{definition}[Level Depth of the \heapName{}]
	\begin{equation} \label{def:lambda-def}
	\lambda_{\alpha} \equiv \log^{\alpha\ast} n  
	\end{equation} 
\end{definition} 

\begin{figure}[H]
	\centering
	\includegraphics[scale=0.7]{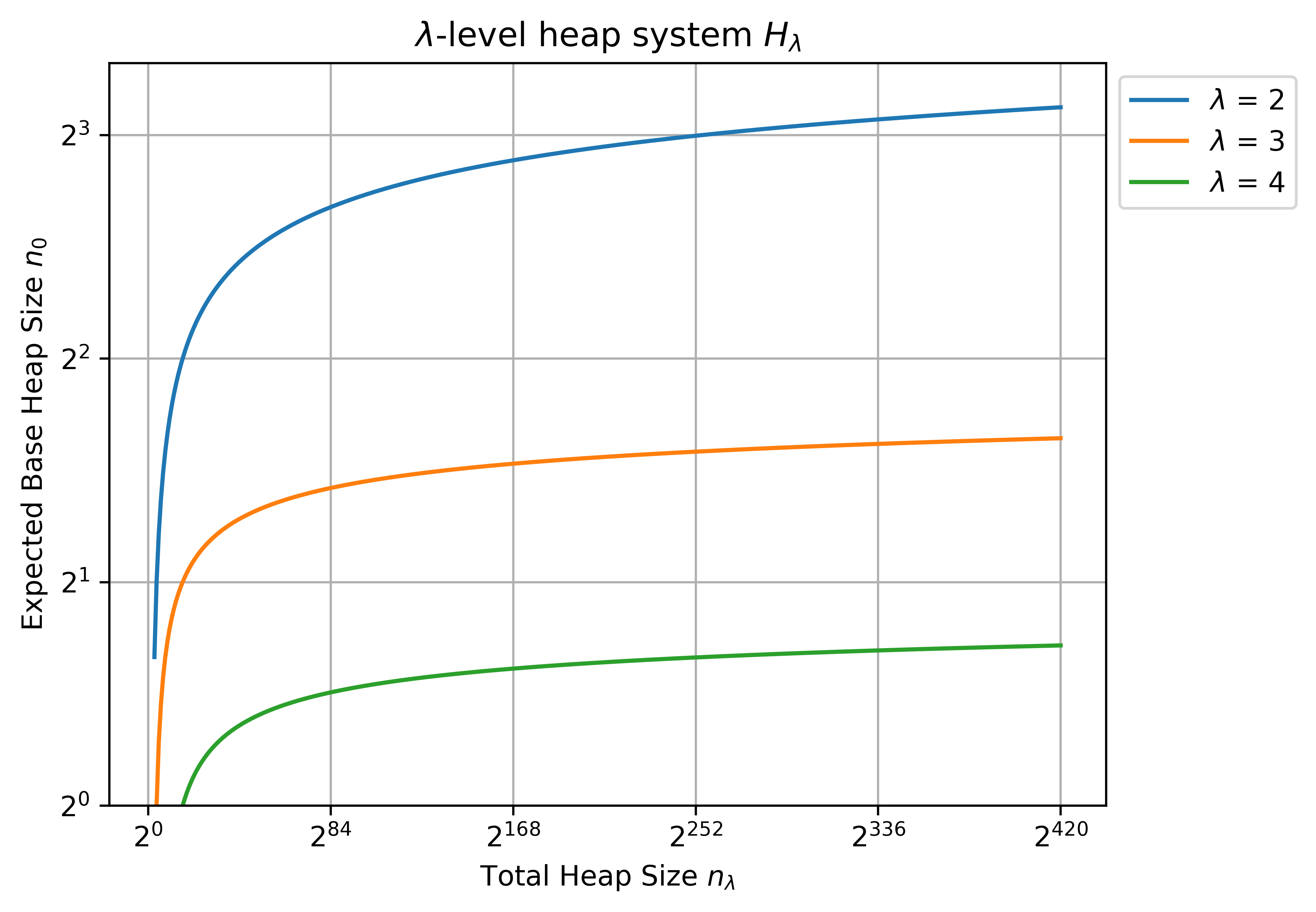}
	\caption {Expected base heap size $n_{0}$ as a function of total heap size $n_{\lambda}$ and level depth $\lambda$. The level depth is defined through the generalized iterated binary logarithm (see Definition \ref{def:log-star}).  \label{fig:small-base-heaps}}
\end{figure}

Since the generalized iterated logarithm $ \lambda_{\alpha}$ terminates strictly faster than the common iterated logarithm $\log^{*} n$ for all problem sizes where $n \le 2^{65536}$ and $\alpha > 1$, the level depth $\lambda_{\alpha>1}$ is less than that for the common iterated logarithm $\lambda_{1}$. In Figure \ref{fig:small-base-heaps}, the expected base heap size $n_{0}$ is depicted as a function of the total heap size $n_{\lambda}$ and for $\lambda_{\alpha}$ with values for $\lambda \in [2\isep4]$. The graph is suggestive of the notion that for total heap sizes $n_{\lambda}$ up to and even more than $2^{420}$ elements, a 3-level heap system would enable robustly maintaining expected base heap sizes $n_{0}$ smaller than or around 4, for which very efficient and highly optimized sort algorithms do exist\cite{Mankowitz}. \\

\subsection{Heap Size Control}\label{heap-size-control}

The heap system $H_{\lambda}$ is designed such that the number of elements maintained on vertically adjacent heap level changes logarithmically (see Definition \ref{def:metaheap-size}). The size of a \textit{metaheap} $k_{\ell}$ and the expected number of maintained elements $n_{\ell-1} = \log n_{\ell}$ for each of its bidirectionally linked children $H_{\ell-1}$ is being determined by the total number of elements $n_{\ell}$ maintained by a heap $H_{\ell}$. Thus, a $\lambda$-level heap system is defined such that at any level $\ell$ with $ \ell \in [1\isep \lambda]$, the elements are assumed to be distributed uniformly over all $k_{\ell}$ heaps of type $H_{\ell-1}$, and each with expected size $n_{\ell-1}$.\\

Formally, we define an implicit heap size control mechanism such that $\forall \ell \in [1\isep\lambda]$, the size of a \textit{metaheap} $k_{\ell}$, i.e. the number of its children, heaps of type $H_{\ell-1}$, is defined by the following relation, 
 
\begin{definition}[Implicit \textit{metaheap} Size Control in a $H_{\lambda}$ System]
	\begin{equation} \label{def:metaheap-size}
		k_{\ell}  := \frac{n_{\ell}}{\log{n_{\ell}}} = \frac{n_{\ell}}{n_{\ell-1}}
	= \frac{\overbrace{\log...\log}^{\lambda-\ell \text{ times}} n_{\lambda}}
	{\underbrace{\log...\log}_{\lambda-\ell+1 \text{ times}} n_{\lambda}} 
	\end{equation}
\end{definition}
\section{Asymptotic Analysis of Time and Space Complexity} \label{analysis}

In the following sections, we define and analyze the runtime characteristics of the recursively defined operations inscribed onto the \heapName{}, an iterated, indexed heap data structure containing elements $[id:p]$, where $id$ is a unique, immutable identifier and $p$ a potentially degenerate, mutable priority of the element. In particular, we will consider mechanisms facilitating the insertion and deletion of elements, priority update operations, such as increase and decrease. The classical operations for obtaining and deleting the element containing the largest priority are also considered. Furthermore, we will evaluate a recursive construction mechanism for the heap system, and its associated space complexity. \\

For the most part, we focus the analysis of all supported operations only on the part of the algorithms acting on the inductive heaps $H_{\ell}$ where $0 < \ell \le \lambda$. The operations on base heaps $H_{0}$ either mirror standard algorithms on array-based binary heaps or are trivial deviations of them. For details on the structure of base ($H_{0}$) and inductive ($H_{\ell}$) heap types, please refer to sections \ref{base-heaps} and \ref{inductive-heaps}, respectively. Recall that all operations on the heap are defined recursively. Thus, the proof structure will follow the typically associated schema of setting up and solving a recurrence relation including a base case ($\ell = 0$) and an inductive step ($ 0 < \ell \le \lambda$). \\

\subsection{Make-Heap} \label{make-heap}

In order to construct a \heapName{} $H_{\lambda}$ with $\lambda$ inductive levels and a base level ($\ell=0$), two distinct types of constructors are employed to generate empty heap instances $H_{\ell} (\ell \in [1\isep \lambda])$ and $H_{0}$, respectively. Although recursively defined, these operations are different in nature from all other   operations supported by the heap system because they are not acting on the system itself but actually creating it. Here, we will focus on and prove by induction Theorem \ref{thm:make-heap} on the computational time complexity of the \texttt{make-heap} operation for inductive levels $H_{\ell}$ where $0 < \ell \le \lambda$, and described by Algorithm \ref{alg:make-heap}. 

\begin{theorem}\label{thm:make-heap}
	Let $H_{\lambda}$ be the \heapName{} constructed to maintain $n_{\lambda}$ elements\footnote[4]{an increase by many orders of magnitude is feasible often for identical values of $\lambda$}, i.e. with $\lambda$ levels of inductive heaps $H_{\ell}$ ($0 < \ell \le \lambda$), and let \texttt{make-heap} be the constructor system of the heap, then its associated time complexity described by Algorithm \ref{alg:make-heap} is $T_{\lambda} = O(n_{\lambda})$.
\end{theorem}

\begin{algorithm}[H]
	\caption{\texttt{make-heap}($\ell$, \textit{n}, \texttt{addressOf}(\textit{bp})=null)} 
	\label{alg:make-heap}	
	\begin{algorithmic}[1] \label{alg:make-heap}
		\REQUIRE See Figure \ref{fig:structure} and Section \ref{inductive-heaps} for nomenclature. \vspace{0.3em} 
		\STATE back\_ptr := bp
		\STATE \textit{heap\_max} := (0, min\_priority value)
		\STATE instantiate \textit{metaheap}, a $k_{\ell}$-sized array of pointers to $H_{\ell-1}$ heaps
		\FOR {$ptr$ in \textit{metaheap}}
		\STATE $ptr$ := \texttt{make-heap}($\ell$-1, $\lceil{\log n}\rceil$, \texttt{addressOf}($ptr$))
		\ENDFOR
		\RETURN \texttt{addressOf}(\underline{this} $H_{\ell}$ instance)
	\end{algorithmic}
\end{algorithm}

We will proceed demonstrating that the recurrence relation describing the running time of the \texttt{make-heap} operation described in Algorithm \ref{alg:make-heap} can be stated as follows,

\begin{equation} \label{eq:make-heap-recurrence}
	T_{\ell} = 
	\begin{cases} 
		c_{0}  & \text{if } \ell = 0\\
		k_{\ell}\:T_{\ell- 1} + k_{\ell} + c_{\ell} &  \text{if } 0 < \ell \leq \lambda
	\end{cases}
\end{equation} 

and whose associated asymptotic solution is $T_{\ell} = O(n_{\ell})$.

\begin{proof} 
	The base level instance $H_{0}$ of the heap system is a simple class containing an empty vector of elements and a small collection of book-keeping variables. The construction thus requires only constant time.\\
	
	On the uppermost level $\lambda$, of which there exists only a single instance, we instantiate a globally used hash-map, the \textit{location\_table} optimized to deal with very roughly $n_{\ell}$ elements. This operation requires expected constant time. \\
	
	 More generally, for heaps $H_{\ell}$ where $0 < \ell \leq \lambda$, the \texttt{make-heap} operation is defined by Algorithm \ref{alg:make-heap}. On lines 1 and 2, we establish a back connection to the parent $H_{\ell+1}$ metaheap, and set initial values for $id$ and $priority$ of the local \textit{\textit{\textit{heap\_max}}} element. Both actions are constant time operations. In line 3, the metaheap for this $H_{\ell}$ instance with $k_{\ell}$ elements is created, taking $O(k_{\ell}) = O(n_{\ell})$ time. Here you may recall definition \ref{def:metaheap-size}, $k_{\ell} = n_{\ell}/\log {n_{\ell}}$. Consecutively, in lines 4 and 5, the $k_{\ell}$ child $H_{\ell-1}$ heaps are created and connected to the $H_{\ell}$ metaheap elements. These $k_{\ell}$ \texttt{make-heap} operations take $O(k_{\ell}T_{\ell - 1})$ time. Finally, the address of the created $H_{\ell}$ instance is returned to its parent in a constant time operation. In summary, we can assert the respective recurrence relation for an $H_{\ell}$ heap to be $T_{\ell} = k_{\ell}T_{\ell - 1} + k_{\ell}$. Thus, the recurrence relation of the entire \texttt{make-heap} constructor system is defined by Eq. \ref{eq:make-heap-recurrence}, and which thus for a \heapName{} $H_{\lambda}$ has the asymptotic solution $T_{\lambda} = O(n_{\lambda})$. 
\end{proof}
\subsection{Find-Max} \label{find-max}

The functionality $\mathtt{find\textrm{-}max()}$ is a part of the core set of heap operations, and by far the simplest operation supported by the \heapName{}. Below we describe its algorithm and determine its time complexity.

\begin{theorem}\label{thm:find-max}
	Let $n_{\lambda}$ be the total number of elements maintained by the \heapName{} $H_{\lambda}$, then the runtime complexity of the operation $\mathtt{find\textrm{-}max()}$ described by Algorithm \ref{alg:find-max} is $O(1)$. 
\end{theorem}

\begin{algorithm}[H]
	\caption{$\mathtt{find\textrm{-}max()}$} \label{alg:find-max}
	\begin{algorithmic}[1] 
		\REQUIRE See Figure \ref{fig:structure} and Section \ref{inductive-heaps} for nomenclature.\vspace{0.3em} 
		\RETURN \textit{\textit{\textit{heap\_max}}}
	\end{algorithmic}
\end{algorithm}

\begin{proof} 
	The base case is trivially true. Any base heap $H_{0}$, an array-based binary heap, stores the maximum priority element in the first position and is thus retrievable in $O(1)$ time. For any heap type $H_{\ell}$, the Algorithm \ref{alg:find-max} simply returns the value of the variable $\textit{\textit{\textit{heap\_max}}}$. Thus, the time complexity of the operation on any level $\ell$ with $ 0 < \ell \le \lambda$ can be formulated using a first-order recurrence relation with constant coefficients exhibiting the solution $T_{\lambda} = O(1)$. 
\end{proof}
\subsection{Delete-Max} \label{delete-max}

The operation $\mathtt{delete\textrm{-}max}$ is also part of the core functionality that is supported by every heap system. We will describe an algorithm for this operation in the \heapName{}, analyze it and prove the following theorem:

\begin{theorem}\label{thm:delete-max}
	Let $n_{\lambda}$ be the total number of elements maintained by the \heapName{} $H_{\lambda}$, and let $\mathtt{delete\textrm{-}max}$ be the supported operation on the heap system described by Algorithm \ref{alg:delete-max}, then its associated time complexity is $T_{\lambda} = O(\log {n_{\lambda}})$.
\end{theorem}

\begin{algorithm}[H]
	\caption{$\mathtt{delete\textrm{-}max()}$}
	\begin{algorithmic}[1] \label{alg:delete-max}
		\REQUIRE See Figure \ref{fig:structure} and Section \ref{inductive-heaps} for nomenclature.\vspace{0.3em}
		\STATE pHeap := \textit{metaheap}[0] 
		\STATE pHeap$\rightarrow$$\mathtt{delete\textrm{-}max()}$
		\STATE restore heap-property in \textit{metaheap} 
		\STATE \textit{\textit{\textit{\textit{heap\_max}}}} := \textit{metaheap}[0]$\rightarrow$$\mathtt{find\textrm{-}max()}$ 	
	\end{algorithmic}
\end{algorithm}

We will proceed verifying the correctness of the following recurrence relation,

\begin{equation} \label{eq:delete-max-recurrence}
T_{\ell} = 
\begin{cases} 
c_{0}  & \text{if }\ell = 0\\
T_{\ell-1} +  \log k_{\ell} + c_{\ell} & \text{if } 0 < \ell \leq \lambda
\end{cases}
\end{equation}

and whose associated asymptotic solution is $T_{\lambda} = O(\log{n_{\lambda}})$.

\begin{proof} 
	On the base heap level ($\ell = 0$), $\mathtt{delete\textrm{-}max}$ is replicating the classical operation of an implicit array-based binary heap, i.e. it restores heap-order in $O(\log n_{0})$ time, and in addition removes the unique identifier and its associated values from the \textit{location\_table} in expected constant time. Since the base heap size $n_{0}$ is defined to be a roughly constant $\delta$, the operation requires expected constant time. \\
	
	On level $\ell$ where $0 < \ell \le \lambda$, lines 1 and 4 are constant time operations. On line 2, we refer work to the $H_{\ell-1}$ heap that lies on the iterated route down towards the particular base heap $H_{0}$ that currently contains the global max-priority element, and thus incurs $T_{\ell-1}$ time.  On line 3, a downward-directed heap-order restoration of the \textit{metaheap} on level $\ell$ is required as result of the heap-order violation induced by the max-priority element. This work requires $O(\log k_{\ell})$ = $O(\log n_{\ell})$ time. 
	Finally, the work $T_{\ell}$ of $\mathtt{delete\textrm{-}max}$ on any level $\ell$ may be summarized by $T_{\ell-1} + \log k_{\ell} + c_{\ell}$, and consequently the runtime complexity in a \heapName{} $H_{\lambda}$ with $n_{\lambda}$ elements, is thus expected $O(\log{n_{\lambda}})$.
\end{proof} 

\subsection{Insert-Delete-Decrease Functionality} \label{Insert-Delete-Decrease}

The operations \texttt{insert}, \texttt{delete} and \texttt{decrease} recursively defined operation on the \heapName{} $H_{\lambda}$ exhibit identical runtime complexities, namely $O(\log^{*} n_{\lambda})$ time, and \textit{metaheap} update probabilities $p = 1/n_{\ell-1}$. They may thus belong to the same category of functions, and are discussed together in the present section.\\

For Algorithms \ref{alg:insert}, \ref{alg:delete} and \ref{alg:decrease}, described in Sections \ref{insert}, \ref{delete}, \ref{decrease} respectively, we show the correctness of the following recursion relation,

\begin{equation} \label{eq:insert-recurrence}
	T_{\ell} = 
	\begin{cases} 
		c_{0}  & \text{if }\ell = 0\\
		T_{\ell-1} + c_{\ell} & \text{if } 0 < \ell \leq \lambda
	\end{cases}
\end{equation}

and whose associated asymptotic solution is $T_{\lambda} = O(\lambda) \le O(\log^{*} n_{\lambda})$.

\begin{theorem}\label{thm:insert}
	Let $n_{\lambda}$ be the total number of elements maintained by the \heapName{} $H_{\lambda}$. The runtime complexity of the operations $\mathtt{insert}$, $\mathtt{delete}$, and $\mathtt{decrease}$ described by Algorithms \ref{alg:insert}, \ref{alg:delete} and \ref{alg:decrease} is expected $O(\lambda) \le O(\log^{*} n_{\lambda})$ time. 
\end{theorem}

\subsubsection{Insert} \label{insert}

\begin{algorithm}[H]
	\caption{$\mathtt{insert(id, p)}$}
	\begin{algorithmic}[1] \label{alg:insert}
		\REQUIRE See Figure \ref{fig:structure} and Section \ref{inductive-heaps} for nomenclature.\vspace{0.3em} 
		\STATE pHeap := uniformly select a random $H_{\ell-1}$ heap via its associated pointer in the \textit{metaheap}
		\STATE metaheap\_update\_required := ( pHeap$\rightarrow$$\mathtt{find\textrm{-}max()}$.p $<$ p ) 
		\STATE \text{\textit{location\_table}[id][$\ell$-1]} := pHeap 
		\STATE pHeap$\rightarrow$$\mathtt{insert(id, p)}$
		\IF {metaheap\_update\_required}
			\STATE restore heap-property in \textit{metaheap}
		    \STATE \textit{\textit{\textit{heap\_max}}} := (\textit{\textit{\textit{heap\_max}}}.p $<$ p) ? (id, p) : \textit{\textit{\textit{heap\_max}}}
		\ENDIF
	\end{algorithmic}
\end{algorithm}

\begin{proof}
	From section \ref{base-heaps}, we know the structural setup of a base heap $H_{0}$. Briefly, on the base heap level ($\ell = 0$), \texttt{insert} is essentially replicating the action of an implicit array-based binary heap, i.e. it inserts an element by retaining heap-order in $O(\log n_{0})$ time. Again, as with the previously described operations above, the base heap size $n_{0}$ is defined to be roughly a small constant $\delta$, and thus the base heap operation may be considered requiring expected constant time.\\
	
	For inductive heaps $H_{\ell}$, i.e. where $0 < \ell \le \lambda$, the \texttt{insert(id, p)} operation is defined by Algorithm \ref{alg:insert} as follows. On line 1, a random $H_{\ell-1}$ heap is selected uniformly via the \textit{metaheap} to distribute insertions across the \textit{metaheap} accordingly. Then, on line 2, it is established in constant time whether the heap-property in the \textit{metaheap} is going to be violated as a result of the element insertion. On line 3, we fill the \text{\textit{location\_table}} of the heap $H_{\ell}$ with required element location information. All three operations require expected constant time. Next, on line 4, we transfer the element insertion action to the corresponding \texttt{insert} functionality on level $\ell-1$, consequently incurring a cost of $T_{\ell-1}$.  Since the expected size of the $H_{\ell-1}$ heap that was earmarked for element insertion is $O(n_{\ell-1}) = O(\log n_{\ell})$, and thus the associated update probability $p_{update}$ of a uniformly random priority being larger than the local max is of reciprocal complexity, in particular, $\frac{1}{ n_{\ell-1}} = \frac{1}{\log n_{\ell}}$, the associated probability of requiring a \textit{metaheap} update-operation with associated cost $O(\log k_{\ell}) = O(\log n_{\ell})$ is thus amortized cost of $O(1)$.\\
 
 	In summary, the work on any level $\ell$  may be summarized by $T_{\ell-1} + c_{\ell}$, and thus the recurrence relation of the \texttt{insert} operation is described by \ref{eq:insert-recurrence}, and the associated time complexity in a \heapName{} $H_{\lambda}$ with $n_{\lambda}$ elements, is thus expected $O(\lambda)$, i.e. $O(\log^{*} n_{\lambda})$ time.
\end{proof}
\subsubsection{Delete}\label{delete}

\begin{algorithm}[H]
	\caption{$\mathtt{delete(id)}$}
	\begin{algorithmic}[1]\label{alg:delete}
		\REQUIRE See Figure \ref{fig:structure} and Section \ref{inductive-heaps} for nomenclature.\vspace{0.3em}
		\STATE pHeap := \textit{location\_table}[id][$\ell$-1]
		\STATE \text{metaheap\_update\_required} := ( pHeap$\rightarrow$$\mathtt{find\textrm{-}max()}$.id $==$ id )
		\STATE pHeap$\rightarrow$$\mathtt{delete(id)}$
		\IF { \text{metaheap\_update\_required}}
		\STATE restore heap-property in \textit{metaheap}
			\STATE \textit{\textit{\textit{heap\_max}}} := (\textit{\textit{\textit{heap\_max}}}.id $==$ id) ? \textit{metaheap}[0]$\rightarrow$$\mathtt{find\textrm{-}max()}$ : \textit{\textit{\textit{heap\_max}}}
		\ENDIF
	\end{algorithmic}
\end{algorithm}

\begin{proof}
	Recall from Section \ref{base-heaps} that the base case of this induction is true. Also at the base level, the unique identifier $id$ and all associated values in the  \textit{location\_table} are removed in expected constant time. Thus, we continue by exploring the situation for inductive heaps, $H_{\ell}$ ($0 < \ell \le \lambda$), and where the $\mathtt{delete}$ operation is defined by Algorithm \ref{alg:delete}.\\
	
	On level $\ell$, using the location-table, we first identify the $H_{\ell-1}$ heap that is maintaining the element with identifier $id$ on level $\ell-1$. Then we proceed to check in $O(1)$ time whether the current max-priority element of the identified $H_{\ell-1}$ heap is the target for deletion. If this is not the case, we merely invoke the level $\ell-1$ instance of the $\mathtt{delete}$ operation, and are done. Otherwise (line 5), we also need to restore the heap property in the level $\ell$ \textit{metaheap} from the point of deletion since the operation potentially induced a violation of the heap order. Since the $H_{\ell-1}$ heap is of expected size $O(n_{\ell-1})$, and any deletion is assumed to be uniformly randomly distributed, this case occurs with  probability $1/n_{\ell-1}$, and requires $O(\log k_{\ell}) = O(\log n_{\ell}) = O(n_{\ell-1})$ time. The amortized cost of this branch is then expected constant time. The total aggregated operational cost for $\mathtt{delete}$ on level $\ell$, is thus the sum of constant time work performed on level $\ell$, and a single operation referred to level $\ell-1$. As a result and including the constant time base case, the total cost of $\mathtt{delete}$ can be expressed by the recurrence relation in Eq. \ref{eq:insert-recurrence}. The total operational cost of $\mathtt{delete}$ may consequently be stated as expected $O(\lambda)$ = $O(\log^{*} n_{\lambda})$ time.	
\end{proof}
\subsubsection{Decrease}\label{decrease}

\begin{algorithm}[H]
	\caption{$\mathtt{decrease(id, p')}$}
	\begin{algorithmic}[1]\label{alg:decrease}
		\REQUIRE See Figure \ref{fig:structure} and Section \ref{inductive-heaps} for nomenclature.\vspace{0.3em}
		\STATE pHeap := \textit{location\_table}[id][$\ell$-1] 
		\STATE \text{metaheap\_update\_required} := ( id $==$ pHeap$\rightarrow$$\mathtt{find\textrm{-}max()}$.id )
		\STATE pHeap$\rightarrow$$\mathtt{decrease(id, p')}$
		\IF { \text{metaheap\_update\_required}}
		\STATE restore heap-property in \textit{metaheap}
		\STATE \textit{\textit{\textit{heap\_max}}} := (\textit{\textit{\textit{heap\_max}}}.id $==$ id) ? \textit{metaheap}[0]$\rightarrow$$\mathtt{find\textrm{-}max()}$ : \textit{\textit{\textit{heap\_max}}}
		\ENDIF
	\end{algorithmic}
\end{algorithm}

\begin{proof} 
	Using the properties of base heaps described in Section \ref{base-heaps}, the base case is true. Thus, we continue by exploring the situation for inductive heaps, $H_{\ell}$ ($0 < \ell \le \lambda$), and where the $\mathtt{decrease(id, p')}$ operation is defined by Algorithm \ref{alg:decrease}.\\
	
	On level $\ell$, using the location-table and the unique identifier $id$, we first determine the particular $H_{\ell-1}$ heap that is maintaining the element on level $\ell-1$. Then we proceed to check in $O(1)$ time whether the current max-priority element of the identified $H_{\ell-1}$ heap is the target of the $\mathtt{decrease}$ operation. If this is not the case, we merely invoke the level $\ell-1$ instance of $\mathtt{decrease}$ on the identified $H_{\ell-1}$ heap, and are done. Otherwise (line 5), we also need to restore the heap property in the level $\ell$ \textit{metaheap} from the point of change. Since the $H_{\ell-1}$ heap is of expected size $O(n_{\ell-1})$, and any priority decrease is assumed to be uniformly randomly distributed, this case occurs with probability $\frac{1}{n_{\ell-1}}$, and requires $O(\log k_{\ell}) = O(\log n_{\ell})$ time, which is equivalent to an amortized cost for this branch of $O(1)$ time. Thus, the aggregated operational cost for $\mathtt{decrease}$ on level $\ell$, is the sum of expected constant time work performed on level $\ell$, and a single operation referred to level $\ell-1$. Finally, the total cost of the operation in a \heapName{} $H_{\lambda}$ can be summarized by the recurrence relation, Eq. \ref{eq:insert-recurrence} and whose operational cost may be stated as expected $O(\lambda)$, i.e. $O(\log^{*} n_{\lambda})$ time.	
\end{proof}
\subsection{Increase}\label{increase}

The heap operations, $\mathtt{increase}$, together with $\mathtt{decrease}$, and $\mathtt{delete}$ constitute the set of operations of the heap system that facilitate the mutability of priority values in elements. All operations rely on a global hash table, the \textit{location\_table}, to retrieve the address of the particular $H_{\ell-1}$ heap that is en route to the element's base heap, whereby the table lookup requires expected constant time. Recall that elements only exist in base heaps, and that is where they are being modified.  \\

\begin{figure}[H]
	\centering
	\includegraphics[scale=0.55]{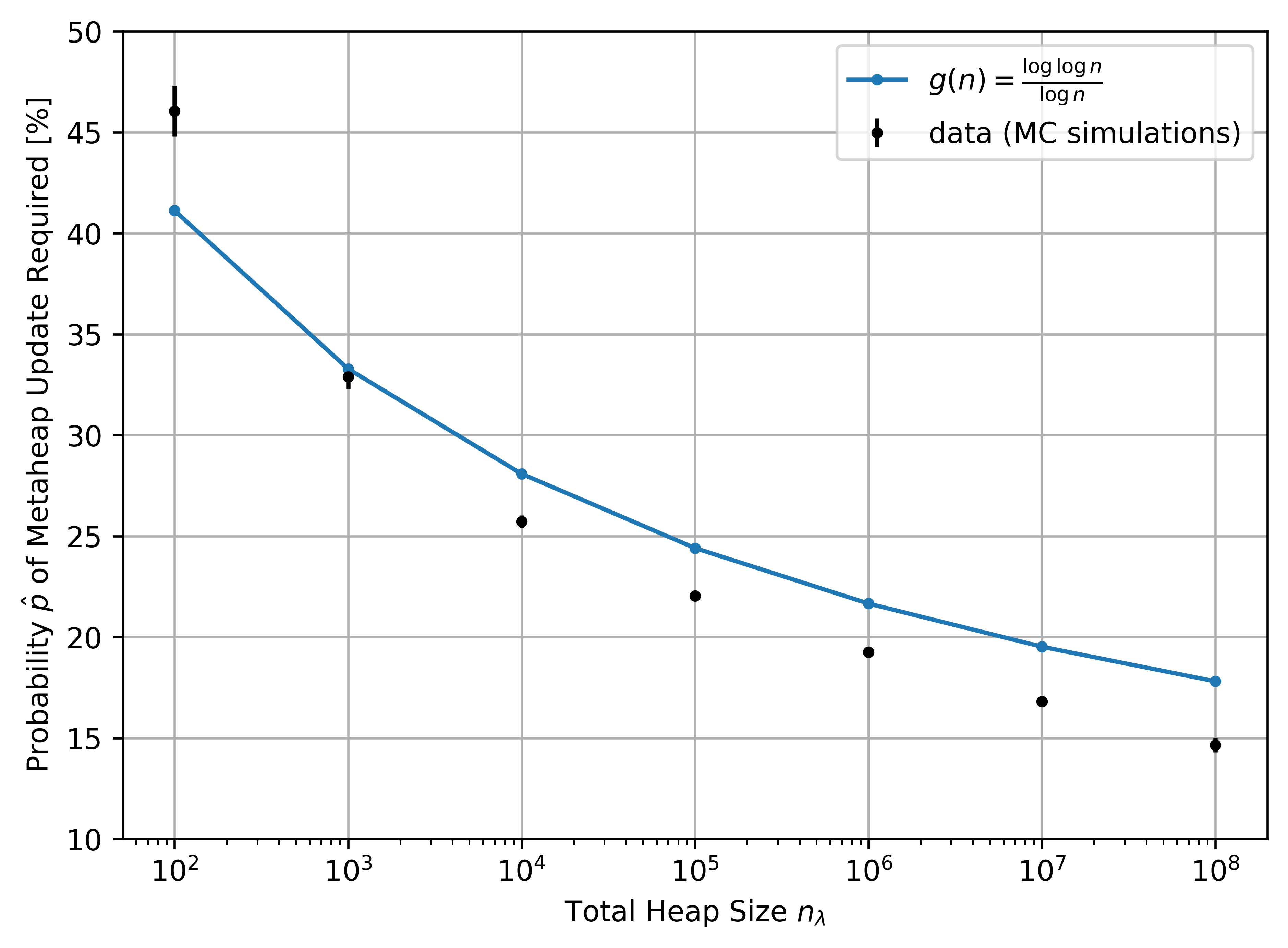}
	\caption {Weak upper-bound estimate of the probability $\hat{p}_{update}$ to require a \textit{metaheap} update as a response to a heap-order violation caused by an $\mathtt{increase}$ operation. Due to limited memory and computing resources, Monte Carlo simulations have only been performed for each order of magnitude of total heap sizes $n_{\lambda}$ in the range [$10^2 \dots 10^8$]. While probably not in the intended optimal operational domain of the \heapName{}, the graph should give a good indication of quantitative behavior. \label{figure:increase-MC}}
\end{figure}

For the operation $\mathtt{increase}$, that is being discussed in this section, the analysis is not fully completed yet. Specifically, a derivation for the expression of the probability of requiring a \textit{metaheap} update, $p_{update}$ (line 4 in Algorithm \ref{alg:increase}), after the priority of a specific element has been increased is still outstanding. To fill this gap in the analysis, and to at least proceed with an estimate of this probability, $\hat{p}_{update}$, Monte Carlo simulations on an accessible range of heap sizes have been performed (Figure \ref{figure:increase-MC}). 

\begin{conj}\label{thm:increase}
	Let $n_{\lambda}$ be the total number of elements maintained by the \heapName{} $H_{\lambda}$. The running time of the operation $\mathtt{increase}$ described by Algorithm \ref{alg:increase} is expected $O(\log\log n_{\lambda})$ time. 
\end{conj}

\begin{algorithm}[H]
	\caption{$\mathtt{increase(id, p')}$}
	\begin{algorithmic}[1]\label{alg:increase}
		\REQUIRE See Figure \ref{fig:structure} and Section \ref{inductive-heaps} for nomenclature.\vspace{0.3em}
		\STATE pHeap := \textit{location\_table}[id][$\ell$-1]
		\STATE \text{metaheap\_update\_required} := ( pHeap$\rightarrow$$\mathtt{find\textrm{-}max()}$.p $< p'$ )
		\STATE pHeap$\rightarrow$$\mathtt{increase(id, p')}$
		\IF { \text{metaheap\_update\_required}}
		\STATE restore heap-property in \textit{metaheap}
			\STATE \textit{\textit{\textit{heap\_max}}} := (\textit{\textit{\textit{heap\_max}}}.p $< p'$) ? $(id, p')$ : \textit{\textit{\textit{heap\_max}}}
		\ENDIF
	\end{algorithmic}
\end{algorithm}

We proceed by conjecturing the correctness of the following recurrence relation,

\begin{equation} \label{eq:increase-recurrence}
	T_{\ell} = 
	\begin{cases} 
		c_{0}  & \text{if }\ell = 0\\
		T_{\ell-1}  + \log\log n_{\ell} + c_{\ell} & \text{if } 0 < \ell \leq \lambda
	\end{cases}
\end{equation}

and whose associated asymptotic solution is $T_{\lambda} = O(\log\log n_{\lambda})$.

\begin{proof}
	 Recall from Section \ref{base-heaps} that the base case of this induction is true. Thus, we continue by exploring the situation for inductive heaps, $H_{\ell}$ ($\ell \in [1\isep \lambda]$), and where the $\mathtt{increase}(id, p)$ operation is defined by Algorithm \ref{alg:increase}.\\
		
	On level $\ell$, and using the \textit{location\_table}, the $H_{\ell-1}$ heap that is maintaining the element with identifier $id$ on level $\ell-1$ is being identified via an expected constant time cost lookup. We proceed to check in $O(1)$ time whether the new priority value p for element id would become the new \textit{\textit{\textit{heap\_max}}} element for its $H_{\ell-1}$ heap. If this was not the case, we simply invoke the level $\ell-1$ operation of $\mathtt{increase}$ on the identified $H_{\ell-1}$ heap, and are done. Otherwise (lines 5 $\&$ 6), since the $H_{\ell-1}$ heap received a new max priority which is now potentially violating the heap order within the level $\ell$ \textit{metaheap}, we also need to restore the heap-property from the position of change up to its root. This update occurs with probability $p_{update}$, for which the estimate $\hat{p}_{update}$ on level $\ell$ based on Monte Carlo simulations appears asymptotically weakly upper-bounded by the function $g(n) = \frac{\log\log n}{\log n} $ (see Figure \ref{figure:increase-MC}). The simulations are only a guide for future work and in order to get a sense of the current expected runtime of the operation without actually deriving an analytical expression. A \textit{metaheap} update operation on line 5 requires $O(\log k_{\ell})$ time, which is equivalent to an amortized cost for this branch of $O(\log\log n_{\ell})$ time.\\ 
	
	Thus, the aggregated operational cost for $\mathtt{increase}(id, p')$ on level $\ell$, is the sum of expected constant work performed on level $\ell$, a single operation referred to level $\ell-1$, and the amortized cost for the \textit{metaheap} update branch, i.e. $O(\log\log n_{\ell})$. As a result, the cost of $\mathtt{increase}$ on level $\ell$ can be expressed by the recurrence relation, $T_{\ell} = T_{\ell-1} + O(\log\log n_{\ell}) + c_{1}$ as stated in Eq. \ref{eq:increase-recurrence}. Thus, upon inclusion of the constant time base case and asserting that $\lambda \ge 2$ for virtually any anticipated use case of this heap data structure, we obtain a computational cost in a \heapName{} $H_{\lambda}$ for this operation of expected $O(\log\log n_{\lambda})$ time.
\end{proof}

\subsection{Space Complexity} \label{space}

In the present section, we will analyze the space requirements of the \heapName{} $H_{\lambda}$, and arrive at the following theorem:

\begin{theorem}\label{thm:space-complexity}
	Let $n_{\lambda}$ be the number of constant-size elements managed by a \heapName{} $H_{\lambda}$, then the space complexity of the data structure $S_{\lambda}$ is $O(n_{\lambda})$.
\end{theorem}

We proceed by verifying the correctness of the following recurrence relation,

\begin{equation} \label{eq:space-recurrence}
S_{\ell} = O(n_{\lambda}) + 
\begin{cases} 
c_{0}  & \text{if } \ell = 0\\
k_{\ell}\:S_{\ell- 1} + c_{\ell} & \text{if }0 < \ell \leq \lambda
\end{cases}
\end{equation} 

and whose associated asymptotic solution is $S_{\lambda} = O(n_{\lambda})$.	
	
\begin{proof}
Firstly, on the uppermost level $\lambda$, there is a single global \textit{location\_table} requiring $O(n_{\lambda})$ space.
Managed elements in the heap system require at least two constant-size data components, namely an identifier \textit{id} and a comparable priority \textit{p}. Thus, in addition to a simple constant-size data type to anchor back to its respective parent $H_{1}$ heap, a base-heap $H_{0}$ contains an array of expected $O(n_{0})$, i.e. constant space.\\

 An inductive heap $H_{\ell}$ contains an array, the metaheap, of expected $O(k_{\ell})$ size, where each of the elements are linked to an $H_{\ell-1}$ heap of size $S_{\ell-1}$. Furthermore, each $H_{\ell}$ heap also contains two  constant-size variables for bookkeeping and to anchor back to its parent $H_{\ell+1}$ heap. Aggregation of all partial space requirements leads to a total cost function for an inductive heap $H_{\ell}$, that may be stated as $S_{\ell} = k_{\ell}\:S_{\ell- 1} + c_{\ell}$. Thus, for an entire heap system $H_{\lambda}$, we establish the total space cost to be correctly represented by the recurrence relation \ref{eq:space-recurrence}. The total space complexity $S_{\lambda}$ of a \heapName{} $H_{\lambda}$ may therefore be stated as $O(n_{\lambda})$.
\end{proof}

\section{Conclusion} \label{conclusion}

The design and analysis of an efficient indexed priority queue with a comprehensive set of operations for priority value management is presented. We show the data structure to exhibit expected $O(\log^{*}{n})$ running times for the operations $\mathtt{insert, delete}$ and $\mathtt{decrease}$, while the expected running time of the $\mathtt{increase}$ operation is stipulated to be weakly upper-bound by a function $h(n) = \log\log{n}$. The time complexity of the heap construction operation as well as the space complexity are demonstrated to be $O(n)$. For $\mathtt{find\textrm{-}max}$ and $\mathtt{delete\textrm{-}max}$, the standard running time complexities of $O(1)$ and $O(\log n)$ are established, respectively. \\

The practicality, robustness and flexibility of the presented data structure are yet to be explored. Despite its appealing computational complexity characteristics, it may well be too involved compared to lean implementations of other highly efficient heap systems in order to be reasonably competitive or useful. However, the outlined setup may have its own distinct benefits for a particular subset of massive graph and/or physical sciences problems if efficient implementations of the data structure can be realized, or necessary improvements can be found. On that note, an index-based as opposed to a pointer-based implementation, especially using contiguous storage for the current destinations of the pointers in the \textit{metaheap}, will very likely exhibit substantially fewer cache misses, and thus proportionally correspond to lower constant factors in the running time characteristic. Thus, it is conceivable that interesting or other useful applications do exist for which this heap system shows great or potentially optimal utility with respect to other heap data structures in certain domains of interest. \\

Finally, with respect to future work on the \heapName{}, establishment of the actual running time of $\mathtt{increase}$ is an outstanding challenge. It is considered interesting to elucidate the concurrency potential of the data structure, as well as the exploration of avenues to create an approximately correct heap system albeit with strict error guarantees in turn for significantly increased speed on a \textit{typical} sequence of operations.

\end{document}